# Charge dynamics in strongly correlated one-dimensional Cu-O chain systems revealed by inelastic X-ray scattering

S. Suga[1], S. Imada[1], A. Higashiya[1,2], A. Shigemoto[1], S. Kasai[1], M. Sing[1], H. Fujiwara[1], A. Sekiyama[1], A. Yamasaki[1], C. Kim[3], T. Nomura[4], J. Igarashi[4,5], M. Yabashi[6] & T. Ishikawa[2,6]

[1]*Graduate School of Engineering Science, Osaka University, Toyonaka, Osaka 560-8531, Japan*
[2]*SPring-8/Riken 1-1-1 Kouto, Mikazuki, Sayo, Hyogo 679-8148, Japan*
[3]*Institute of Physics and Applied Physics, Yonsei University, Seoul 120-749, Korea*
[4]*Synchrotron Radiation Research Center, JAERI, Mikazuki, Sayo, Hyogo 679-5148, Japan*
[5]*Faculty of Science, Ibaraki University, Mito, Ibaraki, 310-8512, Japan*
[6]*SPring-8/JASRI 1-1-1 Kouto, Mikazuki, Sayo, Hyogo 679-5198, Japan*

We report on the Cu 1s resonant inelastic X-ray scattering (RIXS) of Cu-O one-dimensional (1D) strongly correlated insulator systems with contrasting atomic arrangements, namely edge-sharing $CuGeO_3$ and corner-sharing $Sr_2CuO_3$. Owing to good statistics of the high-resolution RIXS data, so far unresolved fine structures are revealed. Detailed photon-energy and momentum dependence of the RIXS spectra in comparison with theoretical calculations has clarified the natures of the low-energy charge excitations and hybridization of the electronic states.

PACS Numbers: 78.70.Ck, 71.20.-b, 71.27.+a, 75.10.Pq

Resonant inelastic X-ray scattering (RIXS) is a powerful tool to probe the momentum dependence of low-energy excitations in solids.[1,2] This technique is intriguing to clarify bulk electronic states of strongly correlated insulators, which are under a keen general interest in decades.[3,4,5] For metallic systems, high-resolution angle-resolved photoemission (ARPES) is promising to detect their occupied states. Nowadays both surface-sensitive low-energy[6] and bulk-sensitive high-energy[7,8] ARPES measurements are feasible. Compared to ARPES, RIXS is really bulk sensitive and applicable to insulators with high resistivity, where the electron correlation is even stronger.[1,2,5,9] However, high energy resolution RIXS is rather demanding because of their poor count rate. For this purpose, high photon flux in a small spot size and a highly efficient analyzer crystal are required.

Here we report photon-energy (hν) and momentum (Δk) dependence of the Cu 1s RIXS with good statistics for two contrasting Cu-O 1D insulating systems $CuGeO_3$[10] and $Sr_2CuO_3$[2] with dominantly divalent Cu. As shown in Fig. 1c, $CuGeO_3$ has a single chain with the edge-sharing $CuO_2$ plane configuration with the Cu-O-Cu angle (θ) of 99°, where the Cu-Cu chain axis is taken as the x-axis and the $CuO_2$ plane corresponds to the x-y plane. The $3d_{xy}$ orbital is unoccupied because it has the highest energy among the whole d orbitals according to an extended d-p model calculation.[11,12] The transfer energy between the neighboring Cu 3d sites via O 2p sites is thought to be very small because of the orthogonality of the Cu $3d_{xy}$ orbitals on the neighboring sites coupled to the O 2p orbitals in the edge-sharing $CuGeO_3$. On the other hand, $Sr_2CuO_3$ has a single Cu-O chain with the corner-sharing configuration as shown in Fig. 2b. The transfer energy is thought to be large in this case, in which the Cu 3d hole is thought to be in the $3d_{x^2-y^2}$ state. Therefore very different behavior of charge dynamics is expected in these systems.

RIXS experiment was performed at BL19LXU of SPring-8 with a 27m long X-ray linear undulator. By use of two channel cut crystal monochromators, the resolution of the incident hν was better than 300 meV. A horizontal focusing was better than 100 μm on the sample. The instrument with 1m Rowland circle was used for the measurement. The total resolution of 400 meV (full width at half maximum) was achieved by using a spherically bent Si(553) analyzer crystal. Transmission (reflection) mode was employed for $CuGeO_3$ ($Sr_2CuO_3$) kept at room temperature. For a thin film $CuGeO_3$ sample, the chain axis was oriented by Laue diffraction. A surface perpendicular to the Cu-O chain was oriented and polished for a $Sr_2CuO_3$ sample. For a dipole excitation, the Cu 1s state is excited to the Cu 4p states, where an electron in a certain occupied state is excited to a certain unoccupied state while emitting the scattered X-rays with the corresponding energy loss. The photon momentum k is large in the X-ray region and the momentum difference Δk between the incident and scattered photons can easily cover few Brillouin zones.

The experimental results of edge-sharing $CuGeO_3$ are shown in Fig.1. The inset in Fig.1c shows the Cu 1s absorption spectrum measured by fluorescence yield. The quadrupolar excitation peak is observed at hν = 8.980 keV, whereas the main absorption band is rather wide. The RIXS spectra are measured for Δk = 3π at three hν of 8.990, 8.995 and 9.000 keV as shown in Figs.1a and b. Three RIXS structures are observed near 6.3, 3.7 and 1.6 eV. It is recognized that the intensity ratio between the structures at 3.7 eV and 6.3 eV is the smallest at hν = 9.000 keV and the structure at 1.6 eV above the smooth tail of the elastic peak is negligible at

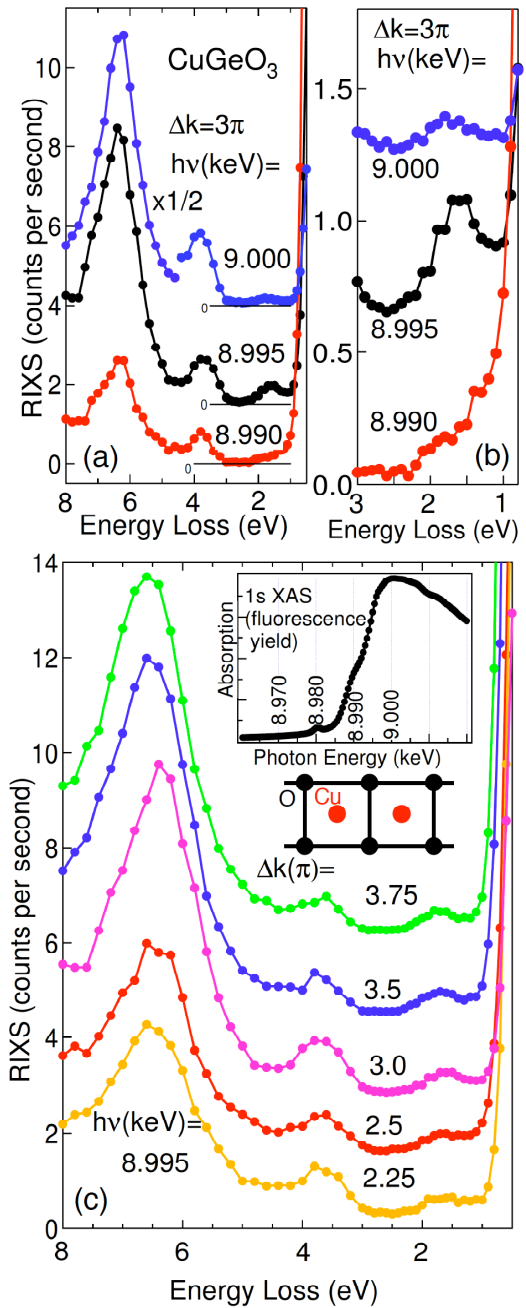

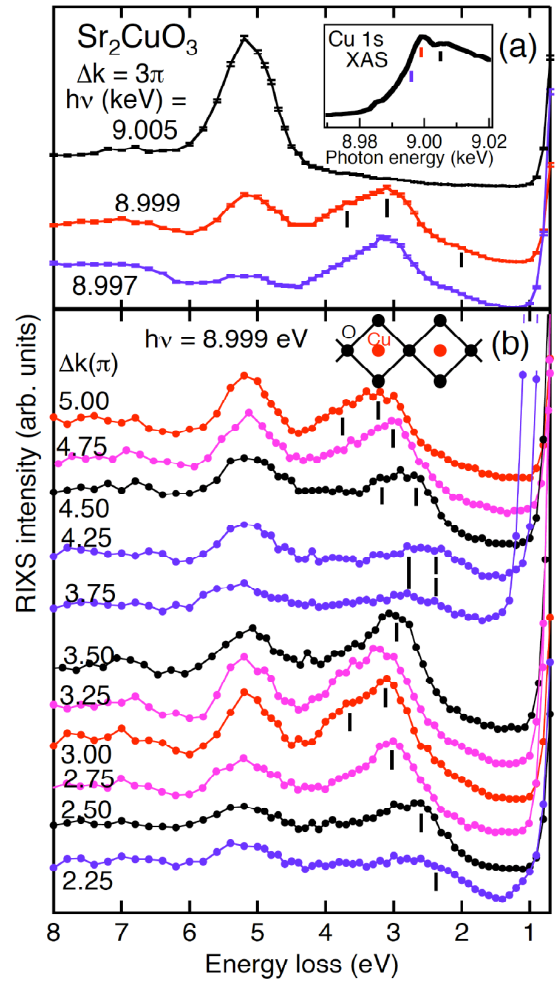

**Fig. 1** (color) RIXS spectra of edge-sharing CuGeO$_3$ at room temperature. The error bar of the intensity is within the size of symbols. a: hν dependence measured at Δk = 3π. b: Same spectra as a but expanded in the low-energy region. c: Δk-resolved RIXS spectra at hν = 8.995 keV. The inset shows the Cu 1s absorption spectrum.

**Fig. 2** (color) RIXS spectra of corner-sharing Sr$_2$CuO$_3$ at room temperature. The error bar is shown, for example, at hν = 9.005 keV. a: hν dependence measured at Δk = 3π. The inset shows the Cu 1s absorption spectrum. b: Δk-resolved RIXS spectra at hν = 8.999 keV. Some representative structures revealed by second energy derivative are indicated by vertical bars for the guide of eye.

hν = 8.990 keV. Then the Δk dependence of the RIXS is measured at different hν. Typical result at hν=8.995 keV (Fig.1c) shows very small dispersion of all RIXS peaks.

The results of Sr$_2$CuO$_3$ are shown in Fig.2. The inset of Fig.2a shows the Cu 1s absorption spectrum. Two peak structures are recognized near 8.999 and 9.005 keV in addition to an absorption hump near 8.985 keV (the quadrupole excitation threshold is near 8.980 keV).

Figure 2a shows clear differences of RIXS at hν = 8.997, 8.999 and 9.005 keV for Δk = 3π. For excitations at 8.997 and 8.999 keV, three broad energy loss structures are observed at around 3.2, 5.2 and ~6.7 eV. One can further recognize a shoulder near 2 eV for hν=8.997 and 8.999 keV and near 3.7 eV for hν = 8.999 keV owing to the good statistics compared with Ref.2. It is remarkable that the 3.2 eV structure is dramatically suppressed at hν = 9.005 keV. Figure 2b shows the Δk dependence of the RIXS at hν = 8.999 keV. A clear Δk dependence is observed for the above mentioned peak located at 3.2 eV at Δk = 3π. Its dispersion is found to have a minimum close to 2.3 eV around Δk = 4π and 2π and a maximum close to 3.2 eV around Δk = 5π and 3π, where shoulders near 3.7 eV and 2.0 eV are also seen. The Δk dependent shift of the structure near 5.2 eV is much less prominent.

We now discuss the observed complex RIXS structures in comparison with theoretical calculations. The extended d-p model calculation by a rigorous numerical diagonalization method for a finite size clusters[11,12] predicts a two-peak structure separated by about 4 eV in the Cu 1s absorption spectrum, reflecting different intermediate states in the RIXS processes. The low energy peak corresponds mainly to the "$|1s^1 3d^{10} \underline{L} 4p^1\rangle$" states ($\underline{L}$ denotes the hole in the O 2p states), where the Cu 3d hole is transferred to the O 2p state in order to reduce the on-site Coulomb repulsive energy between the Cu 1s and 3d holes. The higher energy peak originates mainly from the "$|1s^1 3d^9 4p^1\rangle$" state with the Cu 3d hole on the Cu site. The energy separation between these two peaks depends upon the Cu 3d-O 2p hopping energy and on-site Coulomb energy between the Cu 1s and 3d holes.

We have calculated the electronic structures as well as the charge excitations of the edge- and corner-sharing $CuO_2$ planes within a Hartree-Fock (HF) theory by using a random phase approximation, in which the electron correlation effects are perturbatively taken into account.[13] For $CuGeO_3$, two RIXS peaks located near 3.5 and 6.5 eV with very small dispersions of less than 0.2 eV against $\Delta k$ are predicted, where the spectral weight near 3.5 eV is much smaller than that near 6.5 eV. The 3.5 eV peak corresponds to the excitation from the so-called Zhang-Rice singlet (ZRS), which is made of the Cu 3d hole coupled with the O 2p hole,[14] to the upper Hubbard band (UHB). The energy loss near 6.5 eV is ascribed to the excitation from the bonding state (BS) (in terms of the electron character) between the Cu 3d and O 2p states to the UHB. These structures are in a good agreement with the experimental spectra. Furthermore, the extended d-p model suggests that the intensity ratio between the ZRS→UHB to BS→UHB is enhanced when the low-energy absorption ($|1s^1 3d^{10} \underline{L} 4p^1\rangle$) peak is excited, which is in a qualitative agreement with the hν-dependent experimental result in Fig.1a. On the other hand, the experimentally observed peak at 1.6 eV in the RIXS spectra of $CuGeO_3$ cannot be predicted from both theories. We interpret the 1.6 eV structure as the d-d transition taking place on the same Cu site, as implied from the EELS[15] and soft X-ray O 1s RIXS.[16] Since the $d_{x^2-y^2}$ and $d_{xy}$ orbitals can hybridize via the O 2p orbital in edge-sharing $CuGeO_3$ with θ=99°, the d-d transition takes place between these states. The strong suppression of this peak at hν = 8.990 eV in RIXS is decisively understood because the $|1s^1 3d^{10} \underline{L} 4p^1\rangle$ is dominant at this hν and the d-d transition does not take place in the $3d^{10}$ configuration.

Figure 3 shows the calculated results for $Sr_2CuO_3$ with $t_{x,dp}$ = -1.4 eV, $t_{y,dp}$ = -1.4 eV and $t'_{pp}$ = -0.7 eV and $U_{dd}$=11 eV in the HF theory. Both UHB and ZRS have noticeable dispersions caused by the strong hybridization between the Cu 3d and O 2p states and large transfer energies. The dispersions show the π periodicity reflecting the antiferromagnetic ground state. The calculated RIXS spectra in Fig.3b for typical $\Delta k$ values of $2n\pi$, $(2n+0.5)\pi$ and $(2n+1)\pi$ with integer n show the $2\pi$ periodicity[13] instead of π periodicity, reflecting the partial occupation number of the Cu 3d $_{x^2-y^2}$ electrons for each spin component in the band. The $2\pi$ periodicity is in full agreement with the experimental results. ZRS→UHB excitation is predicted to have a large dispersion and more than two components. It is noticed that its spectral weight shifts toward smaller energies near $\Delta k = 2n\pi$ (n=0, 1, 2 ...) in agreement with the experimentally observed dispersive feature through 2.3-3.2 eV. In this calculation the ZRS→UHB excitation at $\Delta k = (2n+1)\pi$ has a weak low energy shoulder near the energy loss of 2 eV, which is also consistent with the experimental results. Although several arguments have been paid to this threshold structure,[17,18,2] it is demonstrated that this structure is inherent in the ZRS→UHB excitation. The upper and lower BS between the Cu 3d and O 2p states have also

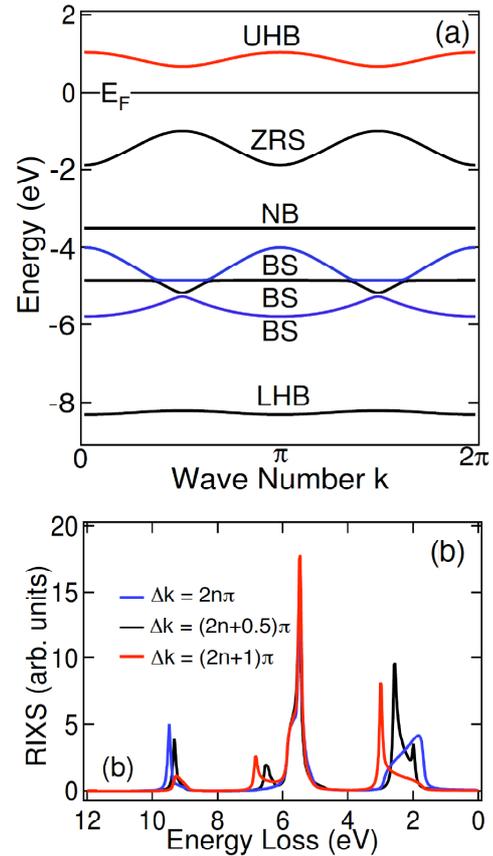

**Fig. 3** (color) a: Electronic structure predicted by the Hartree-Fock calculation (depicted in the electron presentation). Hopping parameters are taken as $t_{x,dp}$ = -1.4 eV, $t_{y,dp}$ = -1.4 eV and $t'_{pp}$ = -0.7 eV with the correlation energy $U_{dd}$ = 11 eV. b: predicted RIXS spectra at three typical $\Delta k = 2n\pi$, $(2n+0.5)\pi$ and $(2n+1)\pi$ with integer n.

noticeable dispersions, whereas the O 2p non-bonding band (NB) at -3.5 and the middle BS at -4.8 eV and the Cu 3d lower Hubbard band (LHB) at -8.3 eV have very small dispersions as predicted in Fig.3a. The smallness of the $\Delta k$ dependence of the 5.2 eV structure in Fig.2b is ascribable to the small dispersion of the middle branch BS→UHB excitation. The excitation from the NB and upper BS states is not strong in RIXS because of the little partial occupation number of the Cu 3d $_{x2-y2}$ electrons. According to this calculation, an additional structure due to the lower BS→UHB excitation is predicted around 6.0-7.0 eV. Such a structure is really experimentally observed in the region between 6 and 8 eV. The intensity of ZRS→UHB excitation is dramatically reduced at h$\nu$ = 9.005 keV. This is because this h$\nu$ corresponds to the intermediate $|1s^13d^94p^1\rangle$ state and further Cu 3d excitation requiring the correlation energy $U_{dd}$ is very unfavorable. The distinct d-d excitation is not observed in corner-sharing $Sr_2CuO_3$ because the hybridization is strong between the Cu $3d_{x2-y2}$ and O 2p states and then the hole is not localized on the Cu site, making the on-site d-d transition not a well defined excitation. On the other hand, the intensity of the ZRS→UHB excitation relative to the BS→UHB excitation is much stronger than $CuGeO_3$, reflecting the easy formation of the ZRS in corner-sharing system. Although the observed structures in the RIXS spectra and their h$\nu$ dependence for $Sr_2CuO_3$ are mostly explained, there is still an unsolved problem with respect to the weak shoulder structure near 3.7 eV for $\Delta k = 3\pi$ (Fig. 2a). The corresponding structure is also traced in smaller energy loss region at $\Delta k$ = 4.50 and 3.75$\pi$ and seems to be related to the main ZRS→UHB branch excitation. A further theoretical study is required to interpret this weak structure.

In conclusion, clear contrasts between the edge-sharing $CuGeO_3$ and corner-sharing $Sr_2CuO_3$ are revealed by virtue of the h$\nu$-dependent and $\Delta k$-resolved RIXS, reflecting the different natures of the electronic states. The high potential of RIXS for the study of strongly correlated insulator systems is thus demonstrated.

The authors acknowledge I. Terasaki, K. Uchinokura and M. Hase for providing $CuGeO_3$ single crystals and K. Tsutsui, T. Tohyama and S. Maekawa for fruitful discussions and D. Miwa for experimental support. This work was partially supported by a Grant-in-Aid for Creative Scientific Research (15GS0213) of Mext, Japan and JASRI (2003B0136-ND3d-np and 2004A0377-ND3d-np).